\documentclass[11pt]{article}

\usepackage{graphicx}
\usepackage{epsfig}
\usepackage{bbm}
\usepackage{amsmath, amssymb}
\usepackage{cite}
\usepackage{subfig}
\usepackage{float}
\usepackage{braket}
\usepackage{resizegather}
\usepackage{multirow}
\usepackage{tikz}
\usepackage{longtable}

\usepackage[margin=2.5cm]{geometry}
\numberwithin{equation}{section}
\usepackage{hyperref}

\usepackage{color}

\newcommand{\IP}{\mathbb{P}}

\newcommand{\cN}{{\cal N}}

\newcommand{\cA}{{\cal A}}
\newcommand{\cB}{{\cal B}}
\newcommand{\cC}{{\cal C}}

\newcommand{\cV}{{\cal V}}

\def\cjn1{{\cA, \cC^*\otimes \wedge^j \cN^*}}
\def\bjn1{{\cA, \cB^*\otimes \wedge^j \cN^*}}
\def\vjn1{{\cA, \cV^*\otimes \wedge^j \cN^*}}
\def\cjn2{{\cA, \cC\otimes \wedge^j \cN^*}}
\def\bjn2{{\cA, \cB\otimes \wedge^j \cN^*}}
\def\vjn2{{\cA, \cV\otimes \wedge^j \cN^*}}

\begin{document}
\begin{flushright}
$ $\\
\end{flushright}
\begin{center}
{
\Large {\bf String Model Building, Reinforcement Learning and Genetic Algorithms}\\[12pt]
\vspace{5mm}
\normalsize
{\bf{Steven Abel}$^{a,}$\footnote{s.a.abel@durham.ac.uk}
},  
{\bf{Andrei Constantin}$^{b,}$\footnote{andrei.constantin@physics.ox.ac.uk}
},  
{\bf{Thomas R.~Harvey}$^{b,}$\footnote{thomas.harvey@physics.ox.ac.uk}
},
{\bf{Andre Lukas}$^{b,}$\footnote{andre.lukas@physics.ox.ac.uk}
},
\bigskip}\\[0pt]
\vspace{0.23cm}
{\it 
{}$^a$IPPP, Durham University, Durham DH1 3LE, UK}\\[2ex]
{{\it 
{}$^b$Rudolf Peierls Centre for Theoretical Physics, University of Oxford\\
Parks Road, Oxford OX1 3PU, UK}\\[6mm]
{\small Based on a talk given by AL at the ``Nankai Symposium on Mathematical Dialogues", 2021.}
}\\[8mm]
\end{center}

\begin{abstract}\noindent
We investigate reinforcement learning and genetic algorithms in the context of heterotic Calabi-Yau models with monad bundles. Both methods are found to be highly efficient in identifying phenomenologically attractive three-family models, in cases where systematic scans are not feasible. For monads on the bi-cubic Calabi-Yau either method facilitates a complete search of the environment and leads to similar sets of previously unknown three-family models.
\end{abstract}

\setcounter{footnote}{0}
\setcounter{tocdepth}{2}

\section{Introduction}
In recent years, advanced computational methods have been used to explore the vast number of models in the string theory landscape. The purpose of this note is to review the recent work of Refs.~\cite{Constantin:2021for,Abel:2021rrj} on applying two methods, reinforcement learning (RL)~\cite{bellman1,bellman1957markovian,Howard,Watkins,Williams,mnih2016asynchronous,sutton2018reinforcement,Ruehle:2020jrk} and genetic algorithms (GAs)~\cite{Holland1975,David1989,Holland1992,Jonesarticle,Reeves2002,Charbonneau:2002,haupt,Michalewicz2004,MCCALL2005205,Whitearticle}, to the problem of finding string theory vacua with certain prescribed properties. More specifically, we will be working within the class of heterotic $E_8\times E_8$ models on Calabi-Yau (CY) three-folds $X$ with vector bundles $V\rightarrow X$ constructed via the monad exact sequence \cite{Distler:1987ee, Kachru:1995em, Anderson:2008ex, Anderson:2008uw, Anderson:2009mh, He:2009wi}
\begin{equation}\label{monad}
 0\longrightarrow V\longrightarrow B\stackrel{f}{\longrightarrow} C\longrightarrow 0\;,\qquad
 B=\bigoplus_{i=1}^{r_B} {\cal O}_X({\bf b}_i)\;,\quad C=\bigoplus_{a=1}^{r_C}{\cal O}_X({\bf c}_a)\; .
 \end{equation}
Here, $B$ and $C$ are each line bundle sums, which can be identified with integer matrices of size $h\times r_B$ and $h\times r_C$, respectively, where $h=h^{1,1}(X)$ is the CYs Picard number. The general problem is to engineer, within this class, models that have a viable phenomenology, and for this study we take that to mean having a standard model spectrum. We will focus on models based on an $SO(10)$ GUT symmetry which requires $V$ to be a bundle with $SU(4)$ structure group and, hence, $r_B-r_C=4$. These GUT theories are broken further to the standard model by a discrete Wilson line on the quotient CY $\hat{X}=X/\Gamma$, where $\Gamma$ is a freely-acting symmetry of $X$. For a fixed CY manifold $X$, the size of this space of monad bundles, allowing for definiteness each integer from ${\bf b}_i$ and ${\bf c}_a$ to assume 10 values, is
\begin{equation}\label{eq:SizeMonadSpace}
10^{h(r_B+r_C-1)}\; ,
\end{equation}
where the $-1$ in the exponent accounts for the condition $c_1(V)=0$ which ensures that the structure group of $V$ is special unitary. This space is sizeable and in particular beyond the scope of a systematic scan even for small values of $h$, $r_B$ and $r_C$. Moreover, standard models within this set are rare and even a large systematic random search may not find even a single example. As we will explain, both RL and GAs can be successfully deployed to explore these model spaces and both methods are able to find standard models efficiently.\\[2mm]
Machine learning techniques were introduced into string theory a few years ago~\cite{HE2017564,Ruehle:2017mzq} with much of the activity since focusing on supervised learning. In this note, we consider reinforcement learning which has been shown to be effective on large environments~\cite{AlphaGoZero} and has previously been applied to string theory in Refs.~\cite{Halverson:2019tkf, Larfors:2020ugo, Krippendorf:2021uxu}. Unlike supervised learning, RL generates the training data during the course of the training. This is done by exploring an environment, guided by a neural network and subject to rewards and penalties. In our context, this environment  consists of the aforementioned monad bundles, specified by the line bundle sums $(B,C)$, on a given CY manifold. Each such monad is assigned a value function $v(B,C)$ which measures the degree of phenomenological viability of the model and reaches its maximum when the model is perfect, that is, if it has all requested properties. The RL reward assigned in an action is computed from the change in this value function.\\[2mm]
Genetic algorithms are not yet widely used in particle and string theory and were introduced to string theory in Ref.~\cite{Abel:2014xta}.  The main idea is to evolve an initial population by crossing individuals subject to their fitness and then adding mutations at a small rate. In our case, the population is of course taken from the same environment of monad bundles that underlies RL and the fitness is equated with the value function $v(B,C)$. This reliance on the same environment allows for a direct comparison of the two methods. \\[2mm]

\section{Environment of monad bundles}
Let us now provide more details of the environment:  as described it consists of monad bundles $(B,C)$ on a given CY manifold $X$, and these are subject to the $E_8$ embedding constraint $c_1(V)=c_1(B)-c_1(C)=0$. To keep the search space finite, the line bundle integers are constrained by
\begin{equation}
 b_{\rm min}\leq b_i^k\leq b_{\rm max}\;,\qquad c_{\rm min}\leq c_a^k\leq c_{\rm max}
\end{equation}
and, for the purpose of GAs, we ensure that the limits satisfy $b_{\rm max}-b_{\rm min}=2^{n_B}-1$ and 
$c_{\rm max}-c_{\rm min}=2^{n_C}-1$, so that every integer in $B$ and $C$ can be binary encoded in $n_B$ and $n_C$ bits, respectively. The resulting bit string of length $\ell=h(r_Bn_B+(r_C-1)n_C)$ will be playing the role of the genotype in the GA implementation.\\[2mm]
As mentioned, the crucial ingredient for both algorithms is the value function $v(B,C)$. This is computed as a sum of contributions that promote models satisfying the following criteria (where most of the required quantities can be explicitly computed from standard methods as, for example, described in Ref.~\cite{Constantin:2021for}, and where we denote $M={\rm max}(|b_{\rm min}|,b_{\rm max},|c_{\rm min}|, c_{\rm max})$): 
\begin{itemize}
 \item The index of the bundle $V$ should be equal to $-3|\Gamma|$, to ensure the ``downstairs" index on the quotient $\hat{X}$ is $-3$, as required for a three-family model. The corresponding contribution to the value function is $-2({\rm ind}(V)+3|\Gamma|)/(hM^3)$.
\item To be able to satisfy the anomaly condition we impose $c_2(TX)-c_2(V)\in\mbox{ Mori cone of }X$, and this translates into a contribution of $\sum_{i=1}^h{\rm min}(c_{2i}(TX)-c_{2i}(V),0)$ to the value function.
\item For $V$ to be a bundle, rather than a sheaf, the degeneracy locus of the monad map $f$ in Eq.~\eqref{monad} needs to be empty, that is, its dimension $d_{\rm deg}$ needs to equal $-1$. Hence, we include the value function contribution $-(d_{\rm deg}+1)$.
\item If the structure group of the bundle is a proper sub-group of $SU(4)$, then the low-energy group enhances. To avoid this we include a value function contribution of $-n_{\rm split}$, where $n_{\rm split}$ is the number of splits of $SU(4)$ into factors.
\item The bundle $V\rightarrow X$ needs to descend to a bundle $\hat{V}\rightarrow\hat{X}$ on the quotient CY, which means $V$ needs to have a $\Gamma$-equivariant structure. This is tested by index divisibility of the line bundles in $B$ and $C$ adding the term $-\sum_{U\subset B,C}{\rm mod}({\rm ind}(U),|\Gamma|)$ to the value function, where $U$ runs over all line bundles or blocks of same line bundles in $B$, $C$.
\item To avoid trivial bundles, we add the contribution $-n_{\rm trivial}$, where $n_{\rm trivial }$ is the number of trivial bundles in $B$ and $C$.
\item A basic stability test on $V$, using Hoppe's criterion, is performed by adding the value contributions $-{\rm max}(0,(h^0(X,B)-h^0(X,C))/(hM^3))$ and $-{\rm max}(0,(h^0(X,B^*)-h^0(X,C^*))/(hM^3))$. The cohomology dimensions required can be efficiently computed from analytical formulae found in Refs.~\cite{Constantin:2018hvl,Brodie:2020wkd,Brodie:2020fiq}.
\end{itemize} 
A state with $v(B,C)=0$ represents a monad bundle leading to an anomaly-free, $SO(10)$ model with the correct number of families which passes non-trivial tests for equivariance and bundle stability. We will refer to such states as perfect states and, in the context of RL, these will be referred to as terminal states.\\[2mm]
Actions $s=(B,C)\mapsto s'=(B',C')$ for RL amount to changing one integer in $B$ by $\pm 1$ and changing an integer in the same row of $C$ by the same amount (so that the condition $c_1(B)=c_1(C)$ is preserved). The reward for such an action is computed from the value function as
\begin{equation}\label{lbreward}
r_{s\mapsto s'} = 
\left\{\begin{array}{ccl}
(v(s')-v(s))^p&\text{if}& v(s')-v(s)>0\\
r_{\rm offset}&\text{if}&v(s')-v(s)\leq 0
\end{array}\right\}
+ r_{\rm step}+r_ {\rm boundary}+r_{\rm terminal}
\end{equation}
where $p\in\mathbb{R}$ is a suitably chosen power and $r_{\rm offset}<0$ is a penalty for decreasing the intrinsic state value. To favour finding terminal states (that is, perfect states) in short episodes a penalty $r_{\rm step}<0$ is added for each step, and actions that lead to the boundary set of the integer space attract a penalty $r_{\rm boundary}<0$. Finally, actions that lead to a perfect model are rewarded with a bonus $r_{\rm terminal}>0$.\\[2mm]
As mentioned, in the context of GAs the same environment is utilised, with the heuristic search being carried out by a population of order 100 individuals, each one being  defined by the  integers $B$ and $C$ of the monad bundles, written into a single bitstring which  plays the role of the genotype. The population is then evolved over generations with a repeated process consisting of three crucial steps: selection for breeding weighted by the individuals' fitnesses, breeding of pairs of individuals by crossing over the genotypes, and post-breeding mutation by randomly flipping a small percentage of the bits in the genotypes. The fitness of each individual after each round of breeding is equated with the value function $v(B,C)$. A crucial aspect in the success or otherwise of the GA is then how the fitness values are translated into the weighting applied to the selection: for this study we employ a rank-weighting selection procedure to select individuals for breeding.\\[2mm]
This environment is considered for two CY manifolds, namely the bi-cubic manifold, which corresponds to a hypersurface of bi-degree $(3,3)$ in $\IP^2\times \IP^2$, and the triple tri-linear manifold corresponding to the intersection of three hypersurfaces of multi-degree $(1,1,1)$ in the coordinates of $\IP^2\times \IP^2\times \IP^2$. These manifolds have $h=2$ and $h=3$, respectively, and their defining equations can be tuned to allow for a freely acting symmetry $\Gamma=\mathbb{Z}_3\times\mathbb{Z}_3$. \\[2mm]
The environment has a redundancy encoded in the permutation group
\begin{equation}\label{groupdeg}
 H\times S_{r_B}\times S_{r_C}\;,
\end{equation}
where $H=S_2$ for the bi-cubic and $H=S_3$ for the triple tri-linear reflects the symmetry of the CY, while $S_{r_B}$ and $S_{r_C}$ arise since permutations of line bundles in $B$ and $C$ do not affect the monad bundle. This redundancy will be kept for the purpose of running RL and GA algorithms and will only be removed from the final list of terminal/perfect states found. In practice, we choose $r_B=6$, $r_C=2$, $b_{\rm min}=-3$, $b_{\rm max}=4$, $c_{\rm min}=0$ and $c_{\rm max}=7$, so that every integer is represented by $n_B=n_C=3$ bits. This leads to large environment sizes
\begin{equation}\label{envsize}
 8^{14}\simeq 4.4\times 10^{12}\quad\mbox{(bi-cubic)}\;,\qquad 8^{21}\simeq 10^{19}\quad\mbox{(triple tri-linear)}\; ,
\end{equation}
which are  certainly not amenable to systematic scanning.\\[2mm]
The environment is realised as a MATHEMATICA package. For the purpose of RL, it is coupled to the policy-based  REINFORCE or actor-critic algorithms, also realised as MATHEMATICA packages. The GA was realised independently as both MATHEMATICA and Python packages as a cross-check, which were then coupled to the same MATHEMATICA environment. As we will see, the different RL algorithms and GA realisations lead to similar results.

\section{RL results}
\begin{figure}[h]
     \centering
    {\includegraphics[width=0.4\textwidth]{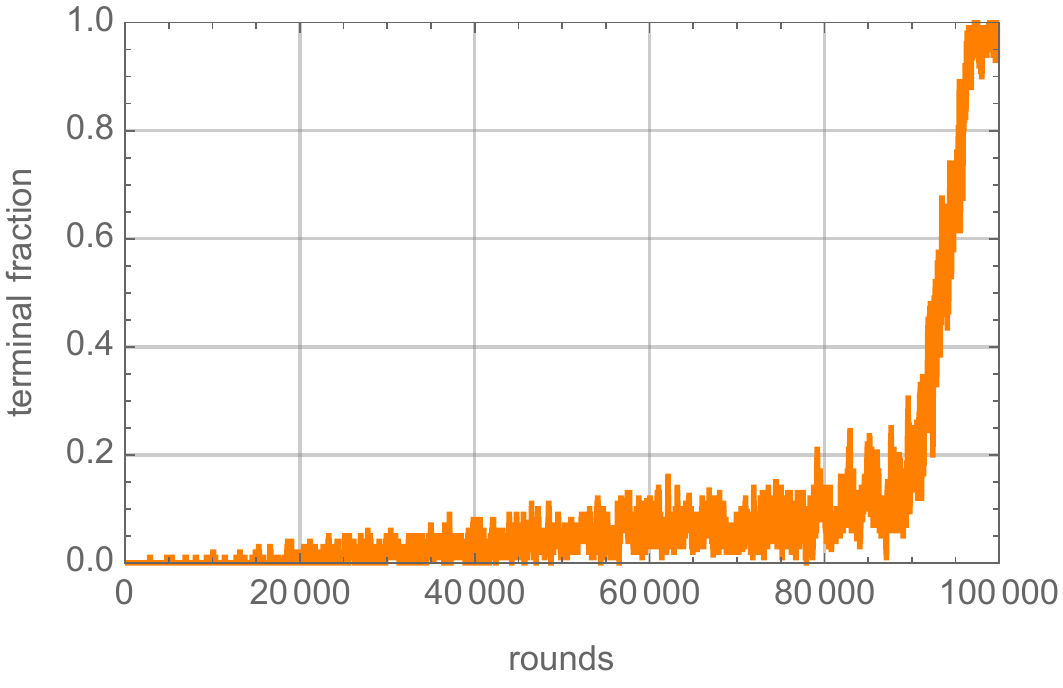}}
    \hspace{21pt}
        {\includegraphics[width=0.4\textwidth]{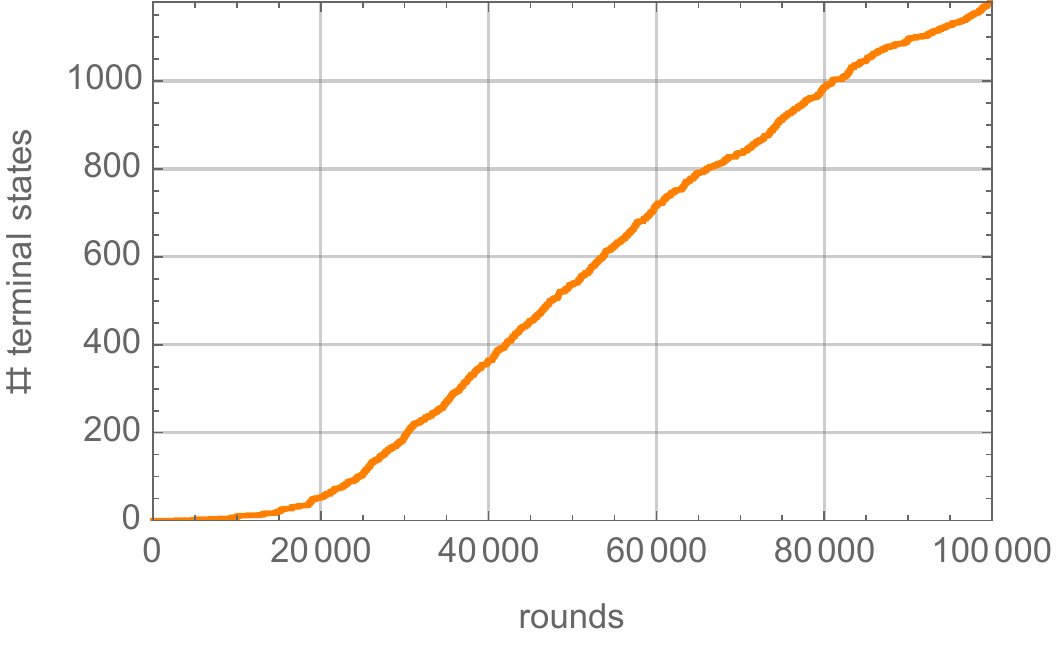}}
       \caption{RL training metrics on the bicubic manifold. After $100\,000$ rounds every episode ends up in a terminal state.}
     \label{fig:RLTraining}
\end{figure}

\begin{figure}[h]
     \centering
      \subfloat[][\centering Number of inequivalent perfect states found as a function of the number of states visited.]{\includegraphics[width=0.47\textwidth]{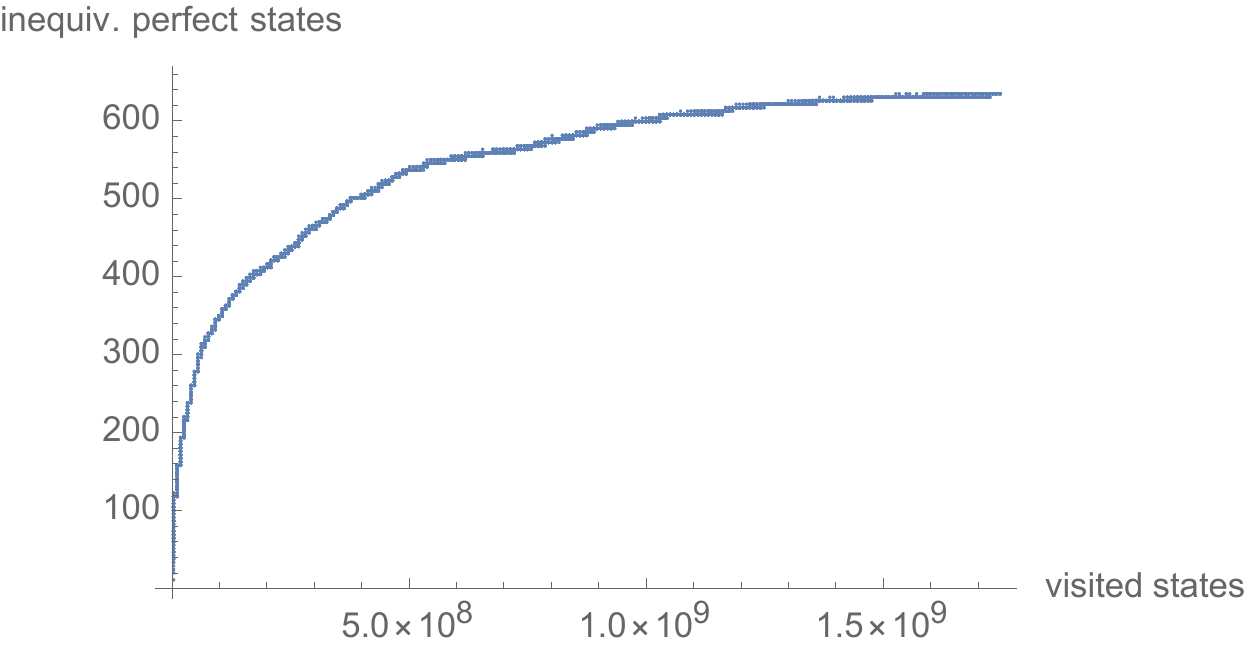}}
     \hspace{21pt}
     \subfloat[][\centering Number of perfect states found as a function of the number of states visited.]{\includegraphics[width=0.47\textwidth]{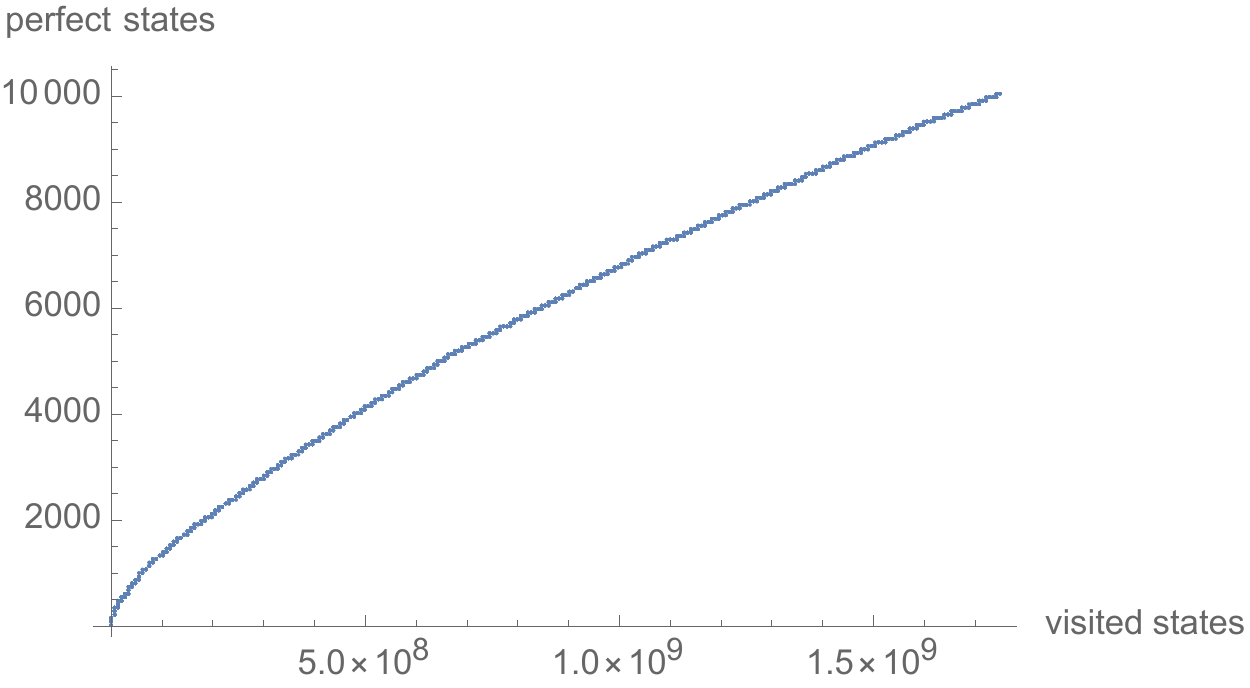}}
     \caption{RL search results on the bicubic. The number of inequivalent perfect states saturates, while the total number of terminal states found is still far from saturation after $1.7\times 10^9$ RL episodes.}
     \label{fig:BicubicRLSaturation1_2}
\end{figure}

We consider policy networks (and value networks for actor critic) which are fully-connected, with typical width $64$ and typical depth $4$, optimised with the ADAM optimiser at a learning rate of $\gamma=0.98$ for REINFORCE and $\gamma=0.8$ for actor critic. For the bi-cubic CY, training such a network with a maximal episode length $64$ leads to a highly successful network after about $100\,000$ training rounds which guides to terminal states from any initial state with an average episode length of about $30$. This only takes a few hours on a single CPU. The corresponding training measurements are shown in Fig.~\eqref{fig:RLTraining}.
\vskip 2mm
\noindent This trained network can be used for a systematic search for perfect models on the bi-cubic by performing many episodes from random initial states guided by the trained policy network. We have carried this out for $1.7\times 10^9$ episodes, taking about $35$ core days and the results are shown in Fig.~\ref{fig:BicubicRLSaturation1_2}.
The saturation in Fig.~\ref{fig:BicubicRLSaturation1_2} (a) indicates that the network has found most of the inequivalent perfect models, $643$ in total. Given the degeneracy of $2!\, 6!\, 2!=2800$ encoded in the group~\eqref{groupdeg} this indicates that there are a few million perfect states and Fig.~\ref{fig:BicubicRLSaturation1_2}(b)  shows that not all of these have yet been found. This is not a problem since the network has found at least one representative in each class and hence all or most of the inequivalent models. Fig.~\ref{fig:BicubicRLSaturation1_2}(b) also shows that the total number of states visited, about $10^9$, is a small fraction of less than $1\%$ of the environment size~\eqref{envsize}.\\[2mm]
For the triple tri-linear manifold RL training is equally successful and leads to a policy network which guides efficiently to terminal states for $100\%$ of episodes from random initial states. However, the number of inequivalent perfect states is much higher than for the bi-cubic and  a complete search has not yet been attempted in this case. Details can be found in Refs.~\cite{Constantin:2021for}.

\section{GA results}
\begin{figure}[h]
     \centering
     \subfloat[][\centering Fitness histogram: number of individuals as a function of generation and fitness.]{\includegraphics[width=0.47\textwidth]{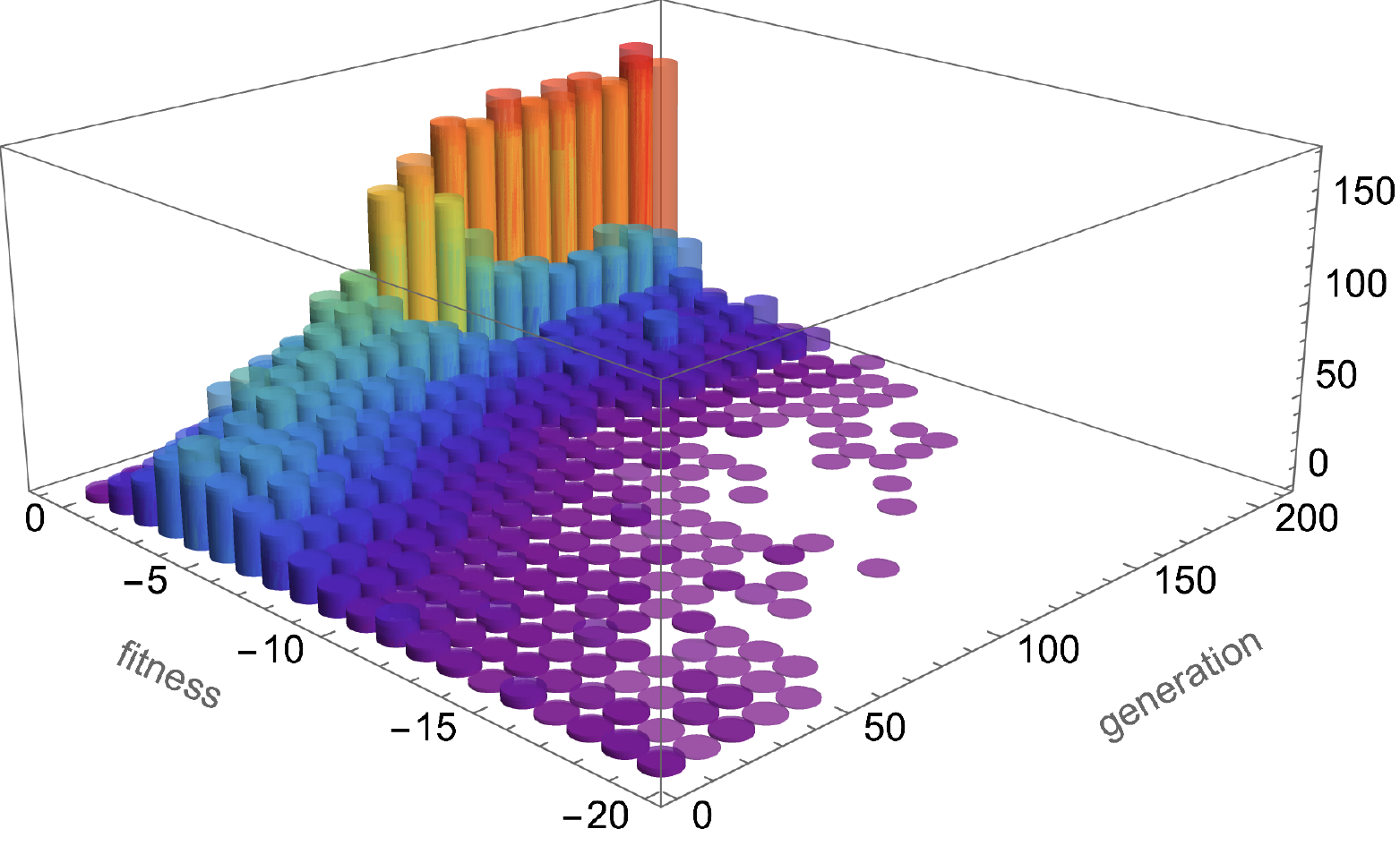}}
     \hspace{21pt}
     \subfloat[][\centering Fraction of perfect models vs generation.]{\includegraphics[width=0.47\textwidth]{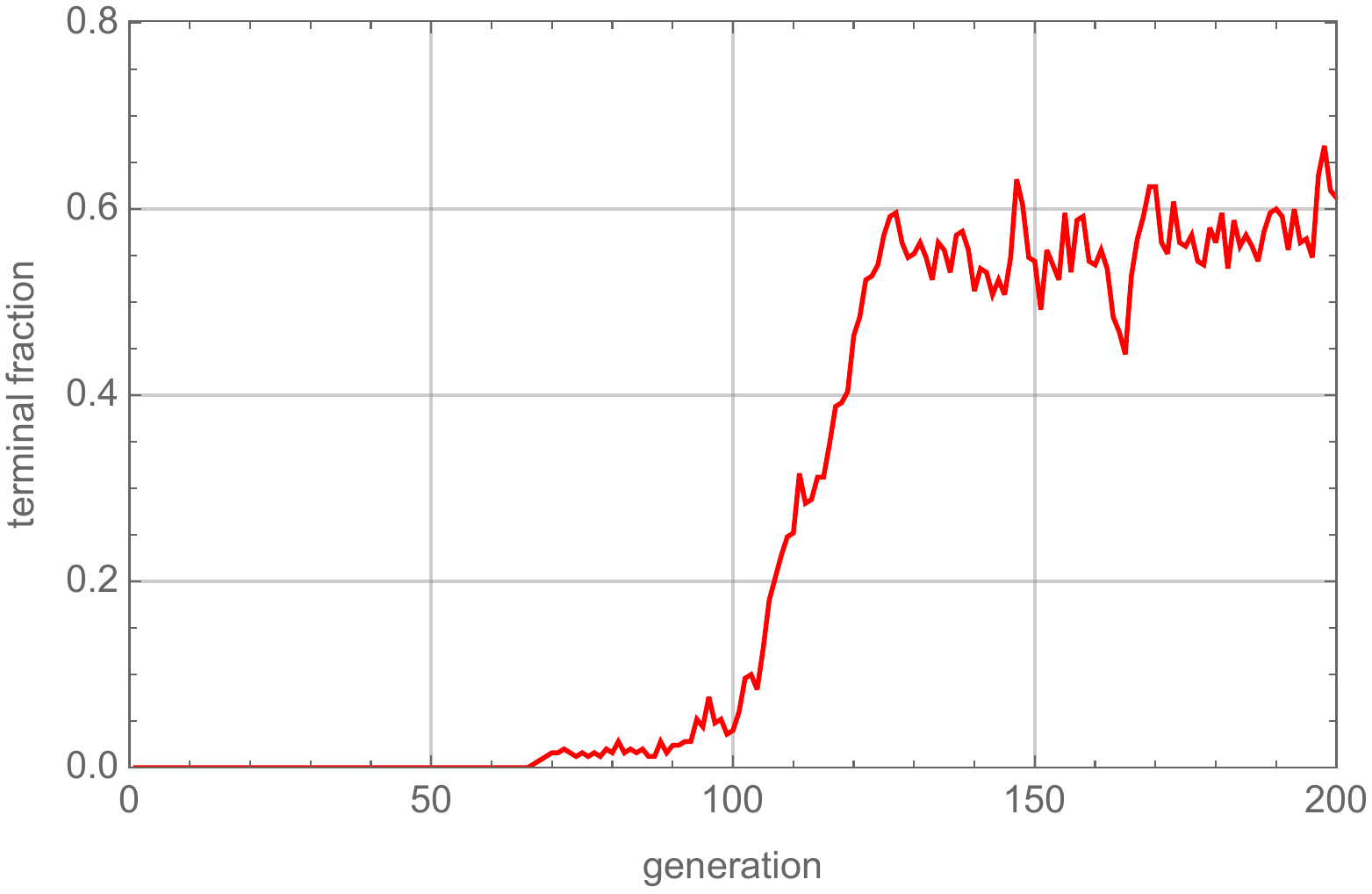}}
     \caption{Typical results for GA evolution on the bi-cubic.}
     \label{figExampleBicubic}
\end{figure}

While the saturation in Fig.~\ref{fig:BicubicRLSaturation1_2}(a) is a good indication that most perfect states have been found by RL it is extremely useful to be able to cross check this with the results found with the GA. For the bi-cubic, we work with a population size $N_{\rm pop}=250$, randomly initialised, and we use a single cut cross-over with rank-weighting selection and a mutation rate of $0.004$. The results of a typical run, which only takes a few minutes on a single CPU, is shown in Fig.~\ref{figExampleBicubic}.
After about $100$ generations roughly $60\%$ of the population consists of perfect states. About $13\,000$ perfect states are contained in the entire evolution, but with only $48$ of them being distinct and $18$ inequivalent. This shows that GAs are impressively efficient in finding a good number of perfect states quickly.\\[2mm]
\begin{figure}[h]
     \centering
      \subfloat[][\centering Number of inequivalent perfect states found as a function of the number of states visited.]{\includegraphics[width=0.47\textwidth]{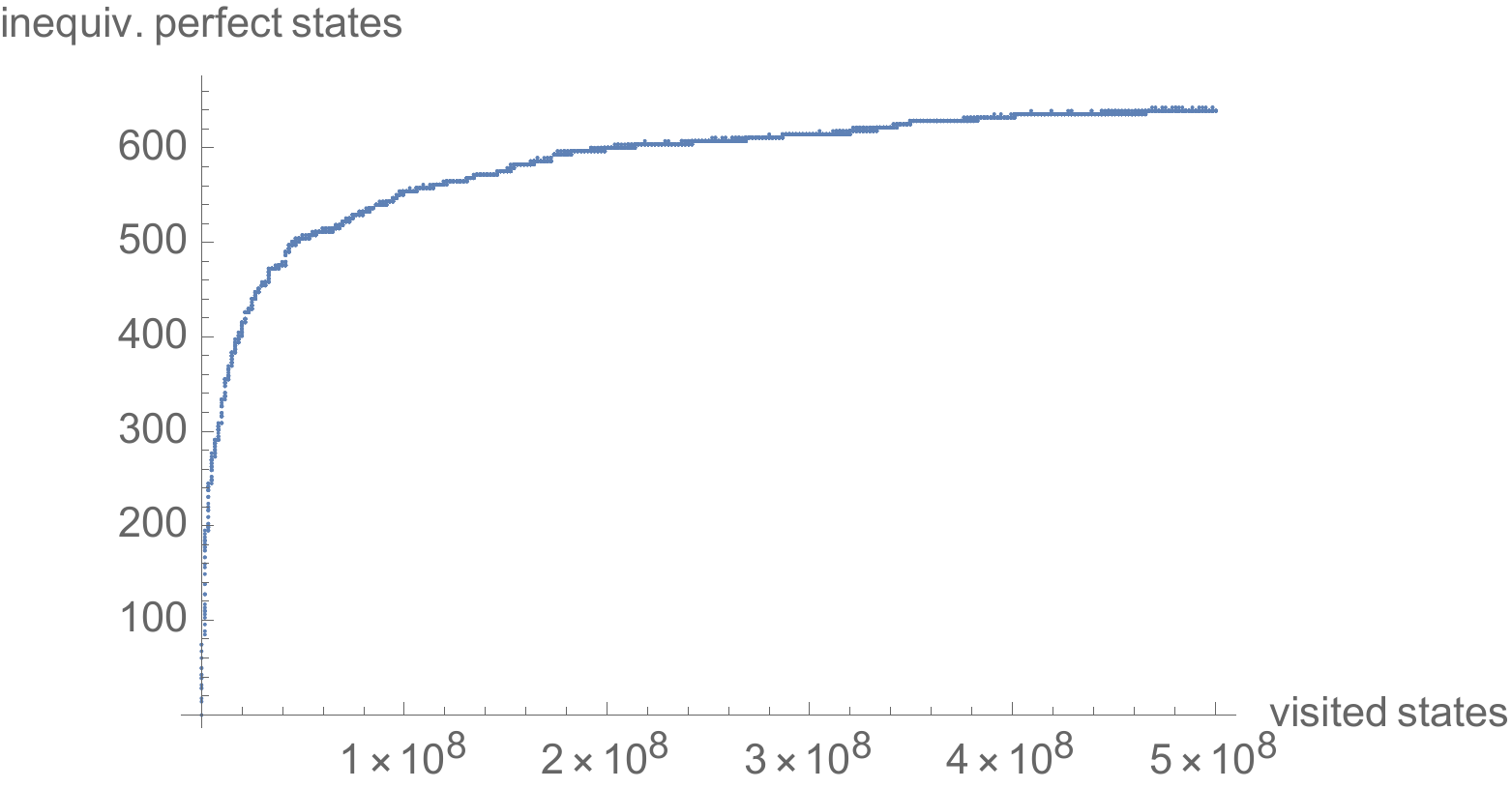}}
     \hspace{21pt}
     \subfloat[][\centering Number of perfect states found as a function of the number of states visited.]{\includegraphics[width=0.47\textwidth]{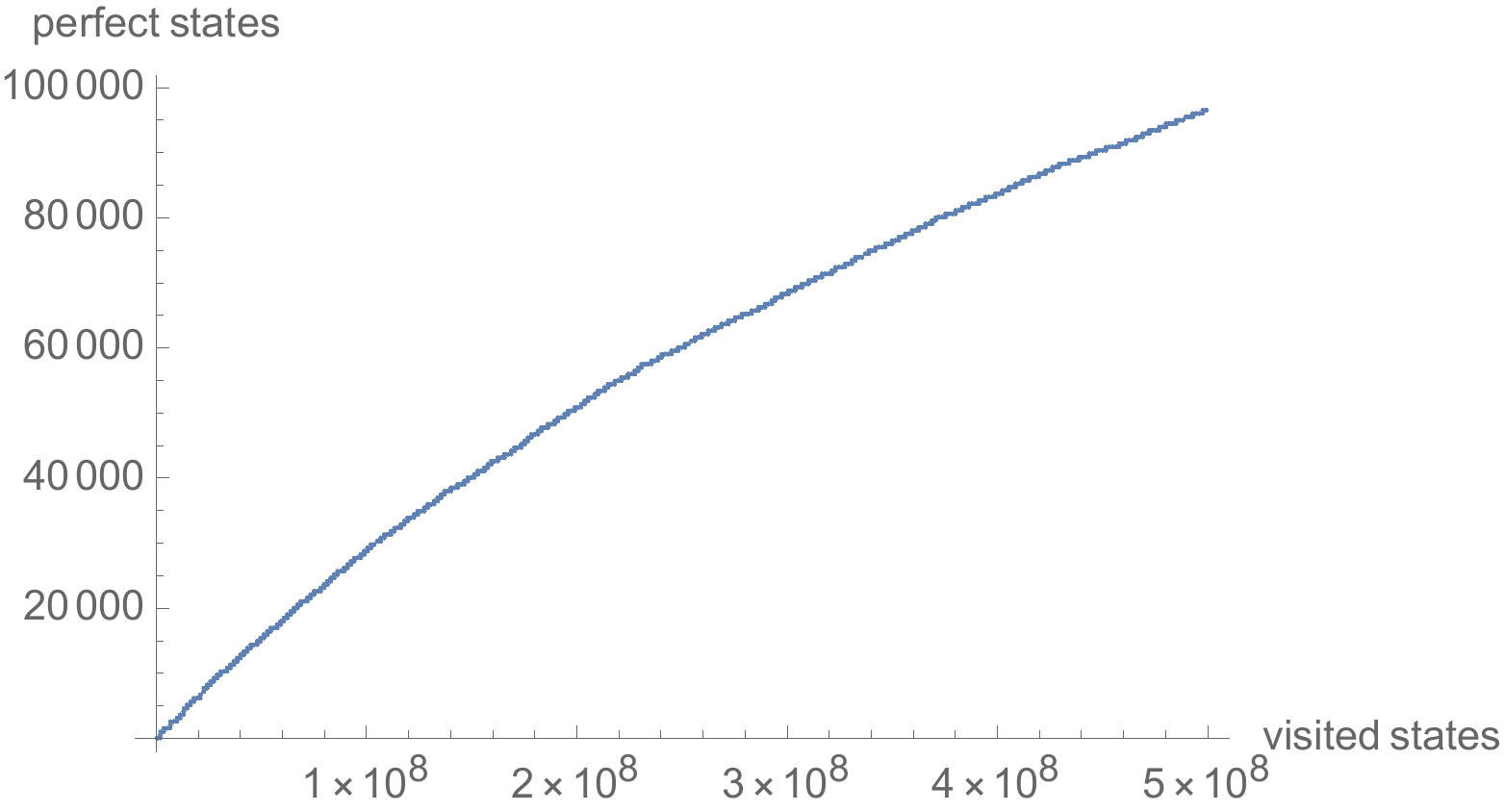}}
     \caption{GA search results on the bicubic. The number of inequivalent perfect states saturates, while the total number of perfect states found is still far from saturation after 10\,000 GA runs.}
     \label{fig:BicubicGASaturation1_2}
\end{figure}
For a more comprehensive search we perform $10\,000$ GA runs with the same parameters as above but with different random initial populations, taking about 10 core days. The result in Fig.~\ref{fig:BicubicGASaturation1_2}(a) shows that the number of inequivalent perfect models saturates and we find a total of $639$ models, about the same as with RL. Moreover, there is significant overlap between the two sets, with only about 50 models in each complement. However, for our configurations, the GA is about $3.5$ times more efficient than RL in reaching this stage. Fig.~\ref{fig:BicubicGASaturation1_2}(b) shows that only a small fraction of the state space in ~\eqref{envsize}, much less than $1\%$, has been visited but many more perfect states have been found than in the RL case (see Fig.~\ref{fig:BicubicRLSaturation1_2}(b)). This indicates that GAs are more prolific at generating equivalent states, a feature presumably related to the cross-over mechanism, which is more likely to lead to permutations of the form~\eqref{groupdeg} than RL actions.\\[2mm]
GAs are equally successful in finding perfect states for monads on the triple tri-linear manifold but due to the significantly larger number of perfect states involved we have refrained from performing a complete search. For details see Ref.~\cite{Abel:2021rrj}.

\section{Conclusion}
We have shown that both reinforcement learning (RL) and genetic algorithms (GAs) are highly efficient in finding perfect (=phenomenologically attractive) models within an environment of heterotic vacua based on monad bundles over Calabi-Yau (CY) three-folds. These environments are typically extremely large, and only a tiny fraction of the models are perfect models, so that systematic scans are out of the question.\\[2mm]
For environments with a relative small number of (inequivalent) perfect states, such as arise on the bi-cubic, either method can be used to find (nearly) all of the perfect states by carrying out a large number of RL episodes or a large number of GA evolutions. The final results show a remarkable overlap between the models found in either case, confirming that a large degree of completeness has indeed been reached, and that the searches are reaching saturation. Most of the $700$ or so perfect models found by either method are new. It turns out, at least for our realisations, GAs are more efficient in reaching this goal. While the final results are comparable, the two methods are somewhat complementary at earlier stages in the search. GAs tend to produce some perfect models rather quickly, within a few minutes computing time, but, in contrast to RL, lead to a large degeneracy under the symmetry~\eqref{groupdeg}. This difference is presumably related to the cross-over method for GAs which is more likely to be compatible with the permutations~\eqref{groupdeg} than RL actions. We think this complementarity is a useful additional tool for performing such heuristic searches. Saturation of a specific search methods might be due to a systematic bias in the procedure that leaves unexplored branches of parameter space. On the other hand, when two complementary methods saturate and yield the same set of perfect models, it is reasonable to conclude that the searches are exhaustive.\\[2mm]
For manifolds $X$ with higher Picard number $h=h^{1,1}(X)$ there exist a significantly larger number of perfect states. This already happens for the triple tri-linear manifolds at $h=3$. Still, both methods are highly efficient in finding perfect models, even on this manifold. However, reaching completeness requires significantly more time. The sensible course of action in this case is to include more refined properties into the definition of the value function, in order to reduce the number of perfect states. This requires performing bundle cohomology calculations for monads which is currently too inefficient to be carried out during RL training or GA evolutions. However, progress on analytic formulae for bundle cohomology~\cite{Constantin:2018hvl,Brodie:2020wkd,Brodie:2020fiq} may well facilitate this soon.\\[2mm]
We hope that more refined definitions of perfect states will, in the future, allow for complete scans of the string landscape for all values of $h$, using RL or GAs. Of course, this assumes running times for these methods only increase polynomially with $h$, an assumption that, albeit plausible from these results, still needs to be verified. It is intriguing that both methods lead to a large fraction of perfect states within an ensemble of episodes or a population. It is tempting to speculate that such searches might be indicative of actual physical processes that occurred in the early universe in which the string landscape was somehow sampled over. Of course we do not know if the evolution of the universe and the selection of the current vacuum was ever subjected to a value or fitness function. However, if it was, then the convergent results of these dissimilar search methods suggests that the resulting distribution of vacua in a multi-verse could be selective and highly non-generic.

\section*{Acknowledgements}
A.~C. is supported by a Stephen Hawking Fellowship, EPSRC grant EP/T016280/1, and T.~R.~H is supported by an STFC studentship. We would also like to thank Sven Krippendorf for suggestion genetic algorithms applied to monad bundles as a possible alternative to reinforcement learning.

\bibliographystyle{utcaps}
\bibliography{bibliography}

\end{document}